# Deterministic control of an antiferromagnetic spin arrangement using ultrafast optical excitation

Y. W. Windsor[1]*, A. Ernst[2], K. Kummer[3], K. Kliemt[4], Ch. Schüßler-Langeheine[5], N. Pontius[5], U. Staub[6], E. V. Chulkov[7], C. Krellner[4], D. V. Vyalikh[7,8], L. Rettig[1]

[1] Department of Physical Chemistry, Fritz Haber Institute of the Max Planck Society, Faradayweg 4-6, 14195 Berlin, Germany

[2] Institute for Theoretical Physics, Johannes Kepler University, Altenberger Strasse 69, 4040 Linz, Austria

[3] ESRF – The European Synchrotron, 71 Avenue des Martyrs, CS 40220, 38043 Grenoble, France

[4] Physikalisches Institut, Johann Wolfgang Goethe-Universität, 60438 Frankfurt am Main, Germany

[5] Institut für Methoden und Instrumentierung der Forschung mit Synchrotronstrahlung, Helmholtz-Zentrum Berlin für Materialien und Energie, Albert-Einstein-Str. 15, 12489 Berlin, Germany

[6] Swiss Light Source, Paul Scherrer Institut, 5232 Villigen PSI, Switzerland

[7] Donostia International Physics Center (DIPC), 20018 Donostia/San Sebastián, Basque Country, Spain

[8] IKERBASQUE, Basque Foundation for Science, 48013, Bilbao, Spain

**A central prospect of antiferromagnetic spintronics is to exploit magnetic properties that are unavailable with ferromagnets. However, this poses the challenge of accessing such properties for readout and control. To this end, light-induced manipulation of the transient ground state, e.g. by changing the magnetic anisotropy potential, opens promising pathways towards ultrafast deterministic control of antiferromagnetism. Here, we use this approach to trigger a *coherent* rotation of the entire long-range antiferromagnetic spin arrangement about a crystalline axis in $GdRh_2Si_2$ and demonstrate *deterministic* control of this rotation upon ultrafast optical excitation. Our observations can be explained by a displacive excitation of the Gd spins' local anisotropy potential by the optical excitation, allowing for a full description of this transient magnetic anisotropy potential.**

## Introduction

Antiferromagnets have attracted great interest in recent years due to their potential to push forward the field of spintronics[1]. A central advantage of antiferromagnetic (AF) over ferromagnetic (FM) spintronics arises from their self-cancelling spin arrangement with zero net moment, rendering them largely insensitive to external fields. Such stability holds great potential for devices (e.g. for digital storage density and longevity), but also poses significant challenges to implement magnetic functionality, requiring new approaches to interact with magnetic order. To this end, a key promise of *antiferromagnetic spintronics* is to exploit and control magnetic properties that are unique to antiferromagnets, such as the ordering wave vector or changes to the spin arrangement itself [2–4]. Another potential benefit of AF spintronics stems from their enhanced magnetic resonance frequencies, promising efficient control of the magnetic state by ultrashort pulses. While in thermal equilibrium a number of approaches for control of AF spintronic systems have been demonstrated[2], more recently also significant progress has been made in manipulating magnetism in antiferromagnetic and ferrimagnetic systems using ultrashort THz[5–8] or visible[9,10] light pulses . However, many of these approaches relied on very specific sample geometries or material properties, e.g. non-centrosymmetric lattice symmetries[2,11].

Another possible route for control of antiferromagnetism using optical stimulation is to utilize the *local* anisotropy of the AF ordered ions to steer the magnetic state of a system. For example, different configurations that are sufficiently close in energy may be susceptible to thermally driven effects on the local (single-ion) magnetic anisotropy. A sudden rise in heat induced by a femtosecond laser excitation can abruptly change this anisotropy, to which the long-range AF order would then respond by realigning according to a new easy axis. In this work, we demonstrate deterministic control of this effect in the prototypical A-type antiferromagnet $GdRh_2Si_2$. The system is an exemplary case of a magnetically soft X-Y antiferromagnet[12], due to the low anisotropy of the closed Gd 4f shell, which has no orbital angular momentum (*L=0*). Its weak magnetic anisotropy makes it ideal for our study, as it allows efficient modification of the anisotropy potential of the Gd ions through ultrafast photoexcitation. Following the excitation, we observe a *coherent* and *deterministic* rotation of the entire AF arrangement of Gd spins about a crystalline axis, without loss of the long-range coherence of the AF order. Further analysis of the coherent rotations of the AF structure also allows us to fully determine the local magnetic anisotropy potential.

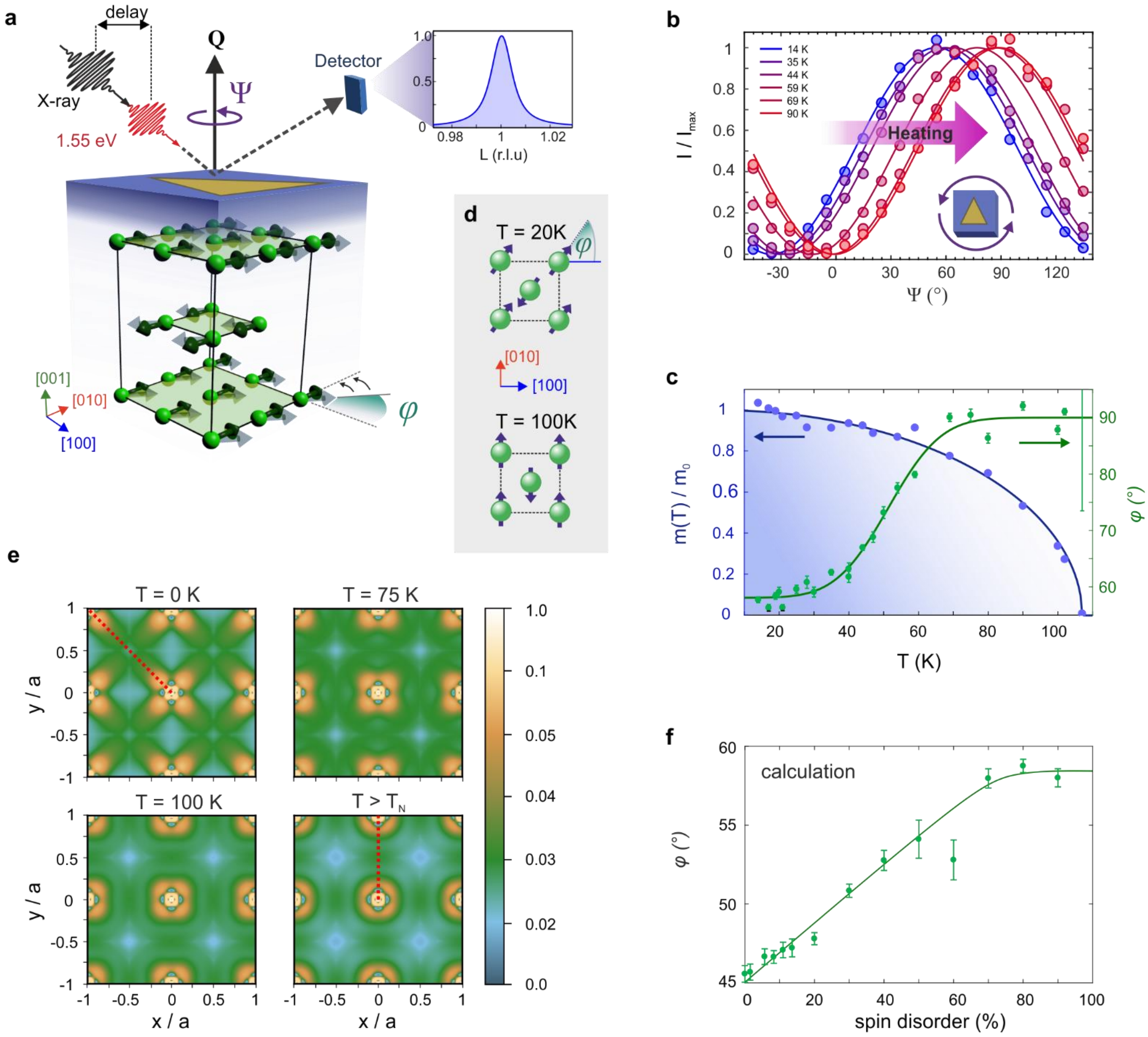

**Figure 1 – Experimental scheme and equilibrium behaviour.** (a) Sketch of experimental scheme and material: the 1.55 eV pump and soft X-ray probe pulses propagate nearly collinearly. They then scatter onto a detector in which only X-rays are detected. Reciprocal space scans produce Bragg peaks, from which integrated intensity is extracted. $\Psi$ indicates azimuthal rotation around $\boldsymbol{Q} = (001)$, which is parallel to the surface normal. A triangle icon is sketched on the sample surface for later reference. A sketch of the layered $GdRh_2Si_2$ crystal structure is shown with Rh and Si ions omitted for clarity (see supplementary figure 1). It indicates two AF spin arrangements (solid and transparent arrows), oriented at an angle $\varphi$ relative to the [100] axis in the *ab* plane. (b) Integrated intensity as function of $\Psi$ collected at selected temperatures in equilibrium (no pump pulse). The vertical axis is normalized to emphasize the change in angular phase. Solid lines are fits to Eq. 1. The icon indicates the $\Psi$ rotation of the sample. (c) Equilibrium behaviour of the Gd 4f spin moment $m(T)$ and angle $\varphi(T)$ (shown in Fig. 1a), extracted from data in (b) using Eq. 1. Solid lines represent fits (see methods). (d) Top down view of the spin arrangement of Gd ions in (a), highlighting the spin rotation caused by the change in $\varphi$. (e) Calculated electron density at $E_F$ within the Gd plane, shown for four different levels of spin disorder: 0%, 50%, 70% and 100% (corresponding to $T$ = 0 K, 75 K, 100 K and $> T_N$ ). The [110] and [010] directions are indicated for clarity by dotted lines. (f) Calculated easy axis angle $\varphi$ as function of 4f spin disorder. The line is a guide for the eye.

## Equilibrium Antiferromagnetic Behaviour

Before investigating the response to laser excitation, we begin by characterizing the equilibrium behaviour of the AF order in $GdRh_2Si_2$. We study its magnetic structure, which consists of ferromagnetic Gd layers (*ab* planes) stacked antiparallel to each other along the [001] direction[13] (the tetragonal axis, Fig. 1a), using resonant X-ray diffraction (RXD) [14,15]. Tuning the incoming X-rays to the Gd $M_5$ edge ($3d_{5/2} \rightarrow 4f$ transitions, see supplementary Figure 1) we gain sensitivity *exclusively* to the long-range AF order of the Gd 4*f* spins by studying the (001) magnetic reflection, which is forbidden for diffraction from the crystal lattice. Since we diffract from ordered spins, the scatterers are not isotropic and the orientation of the spin arrangement relative to the light polarization vector can cause the intensity to depend on the azimuthal angle $\Psi$ (Fig. 1a). Detecting the scattered intensity as function of $\Psi$ therefore allows us to completely determine the orientation of the Gd 4*f* spin moments $m$, which are aligned along an easy axis within the *ab* plane, at an angle $\varphi$ relative to the [100] direction (Fig. 1d). The normalized diffraction intensity of the (001) reflection at various temperatures is shown in Fig. 1b. It exhibits a remarkable shift of the cosine-shaped function with temperature, indicating a change of the magnetic easy axis. Our data can be well described by a magnetic structure factor that depends on $m$ and $\varphi$ as: [16,17]

$$I(x,\Psi) \propto \left|m(x)\cos\big(\Psi - \varphi(x)\big)\right|^2 \qquad x = t \text{ or } T. \tag{1}$$

Here $m$ and $\varphi$ are functions of either temperature ($T$) or pump-probe delay time ($t$, see following). Using Eq. 1 we extract the temperature-dependent behaviour of $m(T)$ and $\varphi(T)$ from the data in Fig. 1b and find that $m(T)$ fits well to a mean-field behaviour for S=7/2 (see methods), and that $\varphi(T)$ changes significantly in the range 30 K – 70 K (Fig. 1c). This corresponds to a gradual rotation of the aforementioned easy axis in the direction from ~[110] towards [010] upon heating (see Fig. 1d). We emphasize that this a *collective rotation of the entire AF spin structure about the [001] axis* (i.e. the material remains AF with vanishing net moment).

To understand this effect, we consider the possible contributions to magnetic anisotropy in $GdRh_2Si_2$. Because the Gd 4*f* shell is spherical ($L = 0$), its anisotropy terms are typically very weak, and only two contributions can occur: from on-site exchange coupling to the conduction electrons which *do* experience magnetic anisotropy (and mediate the RKKY interaction), and from dipolar interactions [18,19]. The dipolar contribution is not expected to contribute to the rotation effect (it itself depends on the spins' orientation), leaving the conduction electrons. Therefore, to gain insight on the observed rotation, the electron density at the Fermi level was calculated for several temperatures. The effect of temperature was simulated as a spin disorder using the disordered local moment approximation (DLM, see methods). Fig 1.e presents the real-space distribution of the electron density within a Gd plane, in which the Gd 5*d* orbital character is dominant. Four different levels of spin disorder are shown, representing different temperatures. For low spin disorder (low temperature) strong hybridization between the Gd 5d states and the Si *sp* bands causes the 5*d* orbitals to extend along the [110] direction (Si ions are the closest to Gd ions). Upon increasing spin disorder (heating), the extent of the 5*d* orbitals along [110] shrinks, causing a reduction in hybridization along [110] in favour of [010] (indicated by dotted lines in Fig 1e). This change in 5*d* anisotropy is imprinted via the strong exchange coupling to the 4*f* states, directly influencing their anisotropy by shifting the overall easy axis angle $\varphi$ in the direction from [110] to [010] (and symmetry equivalents). This explanation for the observed change in the magnetic anisotropy is also corroborated by the calculated easy axis angle shown for several spin disorders in Fig. 1f. It qualitatively reproduces the experimental trend of Fig. 1c, demonstrating the validity of our approach (note that $\varphi$ in Fig. 1f is plotted against spin disorder, in contrast to temperature in Fig. 1c). The rotation effect is also reflected in variations of the Fermi surface in reciprocal space, where degenerate and nearly-degenerate states contribute most strongly to the magnetic anisotropy energy[20]. Upon heating, the direction of highest degeneracy shifts from [110] towards [010] due to an energy shift in the bands near the X point (see supplementary data 1 and supplementary Fig. 3).

## Antiferromagnetic Behaviour upon Ultrafast Excitation

Having established the equilibrium behaviour of the magnetic structure, we now optically excite the system in order to test if we can induce a similar behaviour on ultrafast time scales, as sketched in Fig. 2a. We studied the temporal evolution of the AF structure at a base temperature of 11 K, where $m \approx m_0$ and $\varphi \approx 58°$. We excited the system with 1.55 eV laser pulses, which arrive at delay times $t$ before the femtosecond X-ray probe pulse. The (001) intensity $I(t, \Psi)$ is found to either increase or decrease depending on $\Psi$, but no change is observed in the shape or position of the diffraction peak (Figs. 2b,c). This allows us to track the evolution of $I(t, \Psi)$ by measuring the time-dependent intensity at the maximum of the diffraction peak only. The evolution of $I(t, \Psi)$ with time is strongly fluence dependent, exhibiting both an oscillatory feature and a suppression in magnitude (Fig. 2d).

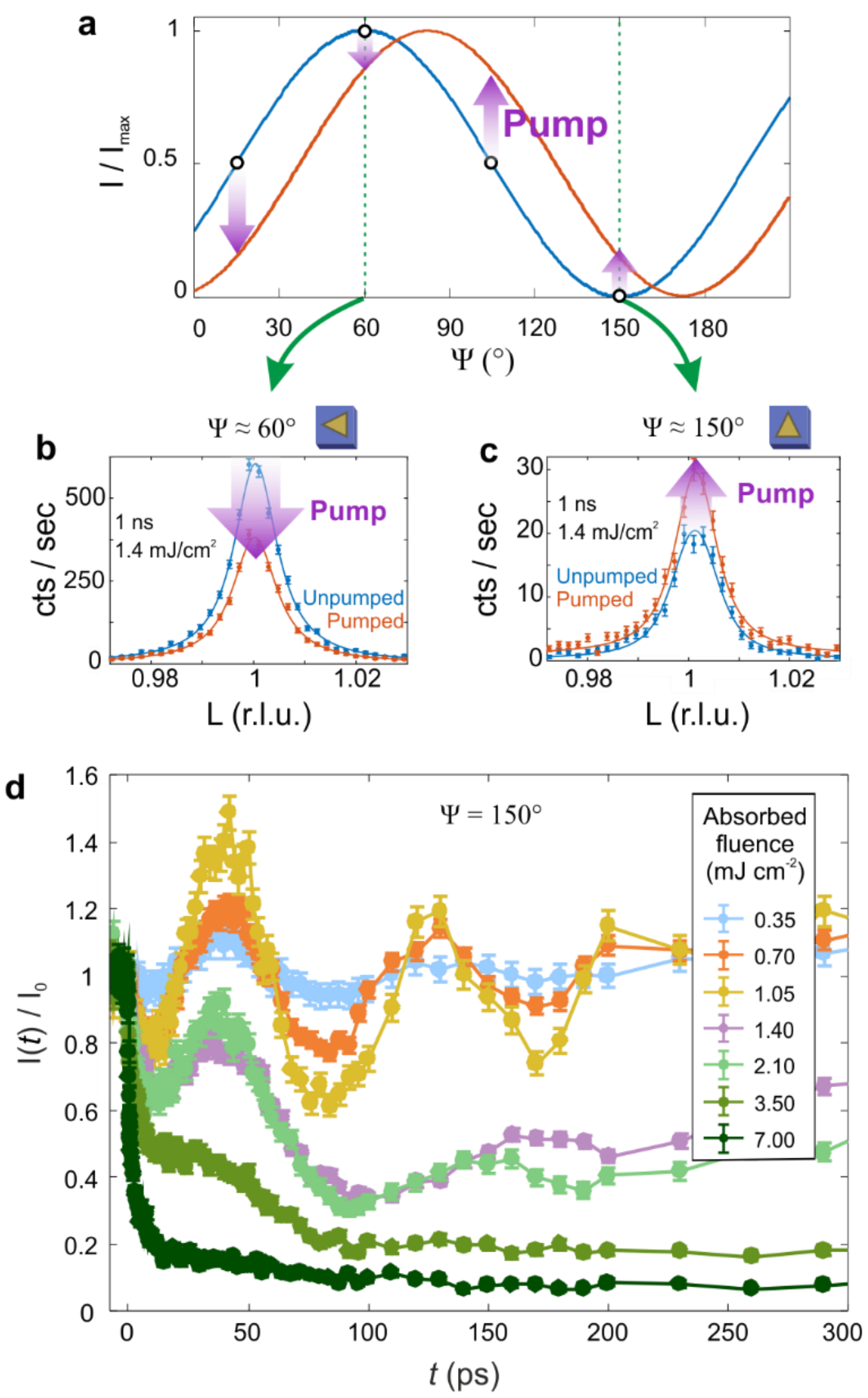

**Figure 2 – Antiferromagnetic response to excitation, time-resolved RXD data.** (a) Sketch of the azimuthal dependence of the diffracted intensity $I(t, \Psi)$ (analogous to Fig. 1b), describing its response to excitation. Circles mark the azimuths at which time-dependent data were recorded. Note that $I(t, \Psi)$ is normalized by $I_{max}$ to highlight the rotational response (this omits the overall intensity reduction due to pump-induced demagnetization). (b) and (c) – reciprocal space scans along $L$ at $\Psi \approx 60°$ and $150°$, respectively, presenting data before and 1 ns after arrival of a 1.4 mJ/cm$^2$ pump pulse. Icons indicate sample orientation. (d) Normalized diffraction intensity of the (001) reflection as a function of pump-probe delay taken at $\Psi \approx 150°$ for several absorbed pump fluences. Error bars reflect photon counting statistics (see methods).

Similar to the equilibrium case, the temporal evolution of $\varphi(t)$ and $m(t)$ can be disentangled using Eq. 1 by measuring $I(t,\Psi)$ for several values of $\Psi$. We measured pump-probe delay scans at four values of $\Psi$, as indicated in Fig. 2a, and used these results to fit Eq. 1 for each value of $t$ (see supplementary note 1). We find that Eq. 1 provides a good description of the transient behaviours of $\varphi(t)$ and $m(t)$ (Fig. 3). $m(t)$ exhibits suppression and recovery dynamics, reminiscent of the ultrafast demagnetization dynamics found in various magnetic systems[21], while $\varphi(t)$ exhibits a combination of a continuous rotation and a superimposed coherent oscillation, corresponding to a collective rotation of the entire AF structure upon photoexcitation (Fig. 2a).

We now discuss the dynamics of $m$ and $\varphi$ in more detail. A hierarchy of the quantities discussed in the next sections is shown in Fig 4a for guidance.

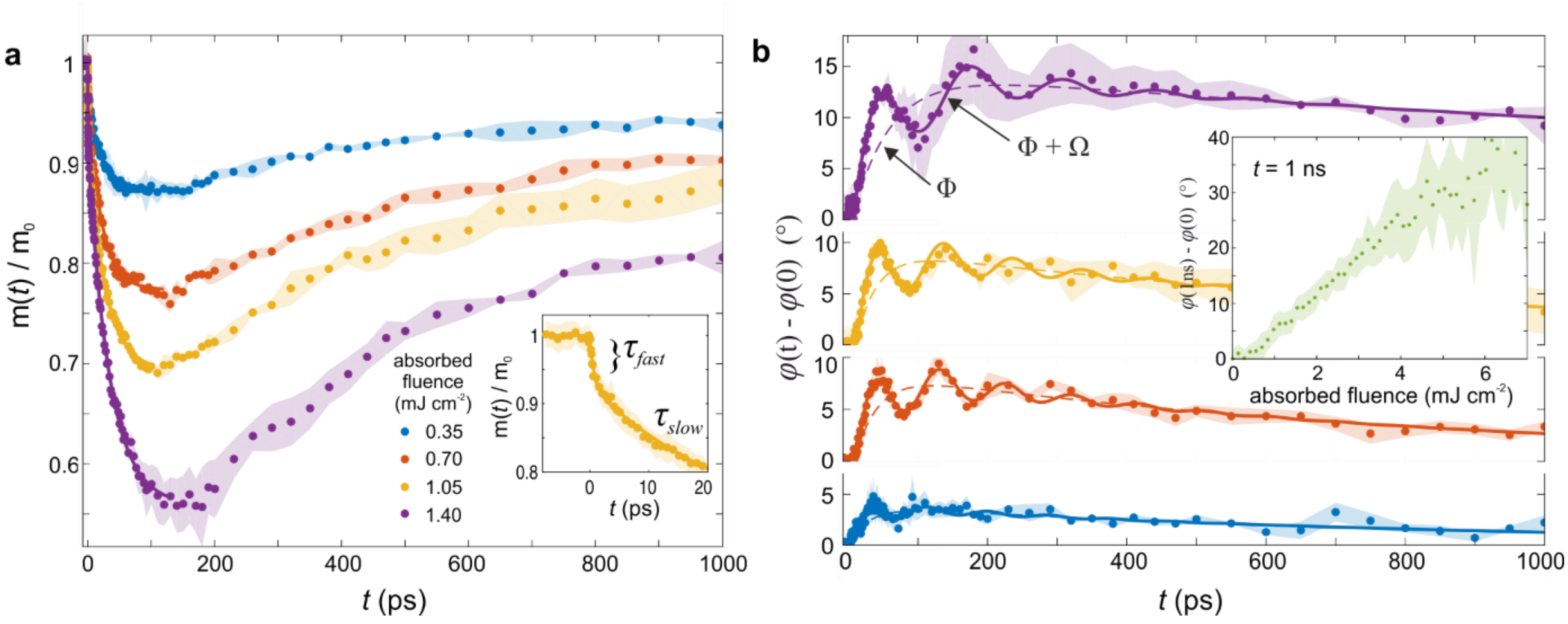

**Figure 3: Ultrafast response of (a) the moment $m$ and (b) the easy axis angle $\varphi$**. For each fluence, the data were extracted from four datasets taken at different sample orientations using Eq. 1. The inset in (a) presents the early delays of the 1.05 mJ/cm$^2$ dataset, highlighting the subpicosecond demagnetization channel that exists for all datasets. Solid lines in (b) are best fits to Eq. 2, and dashed lines are the contribution to the fit from Eq. 3 only (angular shift with no oscillation). The inset in (b) is the angular change at 1ns as function of total absorbed fluence. Shaded areas represent confidence intervals.

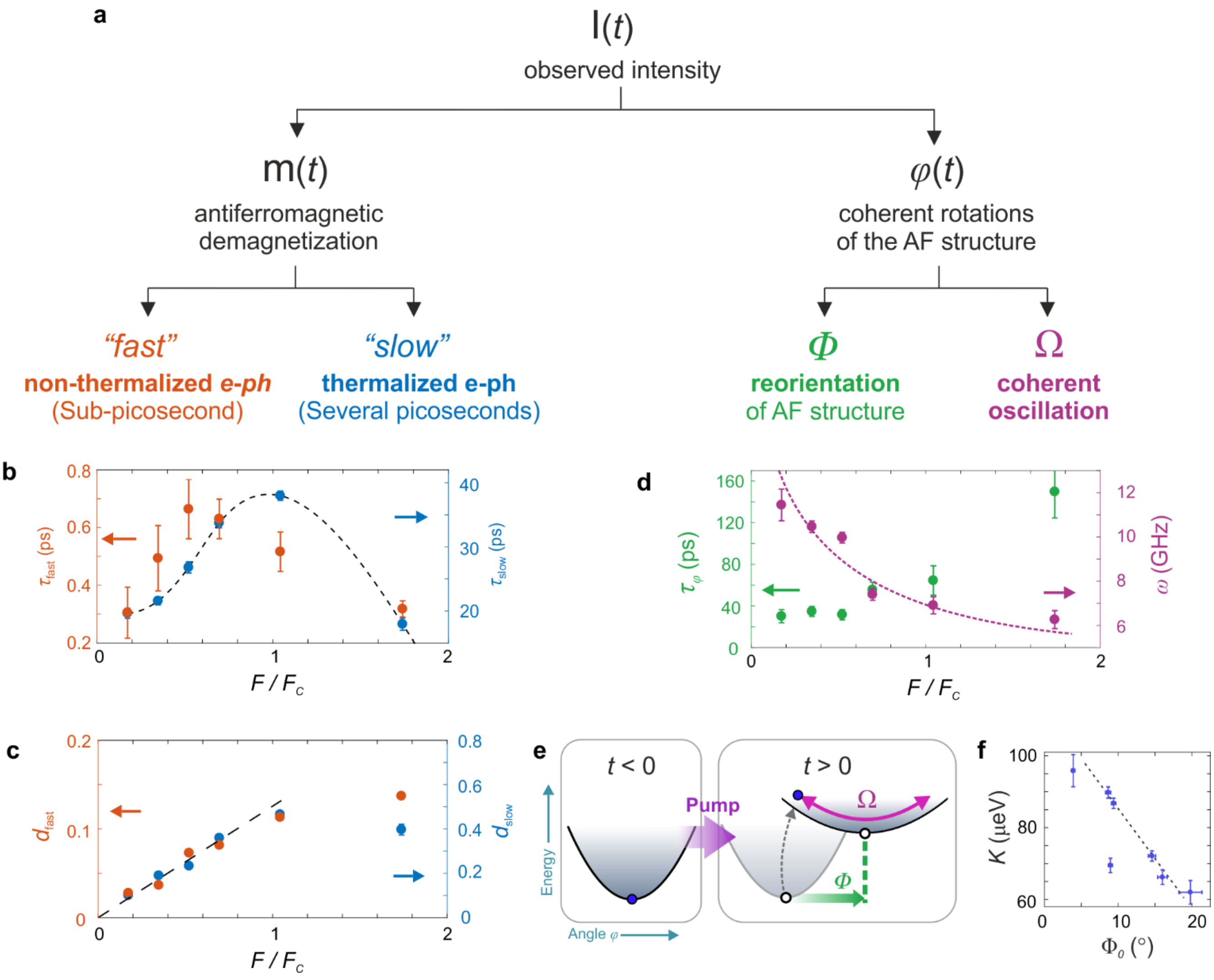

**Figure 4: Contributions to antiferromagnetic dynamics.** (a) Hierarchy of contributions, starting from the observed intensity $I(t)$. (b) and (c) Demagnetization time scales and amplitudes from the "fast" and "slow" channels, as functions of $F/F_C$ ($F_C = 2.01\ mJ/cm^2$). The amplitudes exhibit a linear dependence up to $F/F_C \approx 1$. (d) Fluence dependence of the oscillation frequency $\omega$ and the reorientation time constant $\tau_\varphi$. All lines are guides for the eye. (e) Cartoon describing the AF rotation $\varphi(t)$, highlighting the role of each contribution to the change in the anisotropy potential $U(\varphi)$: $\Phi(t)$ describes a shift of the potential minimum, while $\Omega(t)$ is an oscillation around this new minimum. (f) The transient anisotropy constant $K$ (Eq. 5) as function of the anisotropy rotation amplitude $\Phi_0$ (Eq. 3). The line is a guide to the eye.

The demagnetization of $m(t)$ (Fig. 3a) is well-described by two exponential decay channels (see methods), “fast” and “slow”, with corresponding time constants $\tau$ and demagnetization amplitudes $d$. The *fast* channel occurs on a subpicosecond time scale (300 - 600 fs, see example in the inset of Fig. 3a), while the *slow* channel is $\sim 100$ times slower (20 - 40 ps) and accounts for most of the demagnetization amplitude ($d_{slow} > d_{fast}$, Fig. 4c). Similar bi-exponential demagnetization has been reported for Lanthanide metals, both for AF[22,23] and FM[22,24,25] cases. In all cases the high speed subpicosecond channel accounts for a small fraction of total demagnetization, while the slower channel (i.e. a few ps to tens of ps) accounts for the rest. The appearance of two time scales has been interpreted as one demagnetization process that is initially fast due to the presence of hot electrons which have *not yet thermalized* with the lattice, and then becomes slower once the electrons and phonons have thermalized[26,27]. The time scales we observe (Fig 4b) are in very good agreement with those found for both demagnetization channels of ferromagnetic Gd metal[24,28] and other Gd-based materials[9,25] . Demagnetization amplitudes $d$ exhibit a linear behavior up to a critical fluence $F_C$ (Fig. 4c). In this linear range, the time scales $\tau$ slow down, associated with the rise in susceptibility near the ordering temperature [29]. For the slow channel, this occurs at $F/F_C \approx 1$.

We now discuss the transient rotation dynamics of the AF structure, observed through $\varphi(t)$ (Fig. 3b). The main feature is a gradual reorientation of the AF structure towards higher $\varphi$ values, analogous to the rotation upon heating (dashed lines in Fig. 3b). An additional coherent oscillation of the AF structure is observed superimposed on the gradual rotation. We stress that these behaviours are *collective* and *coherent* motions of the entire AF spin structure, and in analogy to the behaviour upon heating, represent a variation in the ions’ local magnetic anisotropy. The total change in $\varphi$ can be described as

$$\varphi(t) - \varphi(0) = \Phi(t) + \Omega(t)\,. \quad (2)$$

Here, $\Phi$ and $\Omega$ represent the reorientation and the oscillation, respectively (indicated in Fig. 3b). Unlike in the demagnetization data, no subpicosecond dynamics are observed. The gradual reorientation $\Phi(t)$ occurs on timescales $\tau_\varphi$ of tens of ps (Fig. 4d), similar to the values of $\tau_{slow}$. At late delays the reorientation begins to recover on a slower timescale $\tau_{rec}$, such that the whole process can be described as

$$\Phi(\mathrm{t}) = \Phi_0\left(1 - e^{-t/\tau_\varphi}\right)\mathrm{e}^{-t/\tau_{rec}}\,. \quad (3)$$

The amplitude $\Phi_0$ and time constant $\tau_\varphi$ both grow with fluence. Such a dependence is a direct example of a controllable variation in an AF spin arrangement. To demonstrate deterministic control, the inset of Fig. 3b presents the fluence dependence of $\varphi(t) - \varphi(0)$ at a delay of $1\ ns$ (this delay was chosen such that all oscillations are clearly damped). A linear fluence dependence is observed in a wide fluence range, reaching complete rotation of the AF structure towards the [010] direction ($\Delta\varphi > 30$ °) for the highest fluences.

Further insight into the underlying time-dependent magnetic anisotropy potential can be gained from the coherent oscillations $\Omega(t)$. This term exhibits fluence-dependent frequencies $\omega$ of 6-12 of GHz (Fig. 4d), and is damped on time scales $\tau_{damp}$ well-below the recovery of $\Phi(t)$ . It is well described in the form

$$\Omega(\mathrm{t}) = \Omega_0 \cos(\omega t - \zeta)\, e^{-t/\tau_{damp}} \quad . \tag{4}$$

Solid lines in Fig. 3b represent best fits to Eq. 2-4. The observed rotations are reminiscent of displacive excitations of coherent phonons (DECP)[30], in which a step-like change in a vibrational potential triggers coherent phonon oscillations[a]. In analogy, we describe the local in-plane magnetic anisotropy of the Gd spins as a potential $U(\varphi)$ of magnitude $K$, centred around the equilibrium angle $\varphi_0$ as

$$U(\varphi) = \frac{1}{2} K \sin^2(\varphi - \varphi_0)\ . \tag{5}$$

In this picture, the coherent spin oscillations are triggered by the rapid change of the easy axis direction $\varphi_0$ expressed through $\Phi(t)$ in Eq. 3, as depicted in the cartoon in Fig. 4e. Eq. 4 represents a solution of the equation of motion within this transient anisotropy potential, describing an oscillation of the AF structure around the transient easy axis, at a frequency $\omega$ corresponding to the transient value of the magnetic anisotropy constant $K$. The extracted values of $K$ (see methods) are presented in Fig. 4f as a function of $\Phi_0$, the amplitude of the transient easy-axis rotation (Eq. 3). They are remarkably close to those of Gd metal[18], and the corresponding anisotropy fields reach $0.24\ T$, in good agreement with results in Ref.[31].

[a] Displacive excitations are typically described by a phase $\zeta = 0$, accounting for a step-like shift of the potential minimum, compared to the oscillation response time[30]. Here these response times are of similar order (but rotation times $\tau_\varphi$ are still shorter than half the oscillation period, as required for displacive excitations, see Fig. 4d). This more gradual turn-on of the oscillations leads to a finite oscillation phase $\zeta$ (see methods).

Our method provides a complete description of the transient magnetic anisotropy, which exhibits a linear relation $K \propto \Phi_0$ in Fig. 4f, demonstrating that both quantities (easy-axis angle and anisotropy strength) change together upon excitation. Assuming the transient system to be in a thermally equilibrated state, our transient values of $K$ can be interpreted as a measure of the equilibrium anisotropy potential. This assumption is reasonable, considering that the reorientation occurs over tens of ps, which is substantially longer than typical e-ph thermalization time scales in metals of only a few ps[32]. Thus, our results demonstrate not only a method to control the AF spin structure using light, but also to probe the anisotropy potential of an antiferromagnet through a displacive excitation, which can be understood as an antiferromagnetic $\Gamma$-point magnon[33].

In this work, we have demonstrated the optical manipulation of the local (single-ion) magnetic anisotropy via its temperature dependence. Beyond such thermal effects, strong variations in this energy landscape could also be achieved by other means, such as directly tuning magnetic interactions, varying the crystal field, distorting the local environment with an excited phonon mode, and more. Such approaches could also offer faster response times to the excitation. Ultimately the purpose is to embed such functionalities in spintronic devices. For example, in the context of magnetic memory such effects could facilitate ultrafast all-optical control of the pinning direction in write-head spin valves, or enable ultrafast steering of spin current directions in exchange-biased layers. Understanding controllable magnetic properties is challenging with antiferromagnets because of their inaccessibility through conventional means. The insight gained on $U(\varphi)$ through our approach was enabled by the relatively slow GHz frequencies characteristic of our model material system. Antiferromagnets that are candidates for applications typically exhibit THz frequencies[34], which renders similar effects more appealing for future devices, but also more challenging to identify. The effects we present may also be applicable beyond the scope of antiferromagnets, because of the Gd ions' combination of high spin moment and low anisotropy. For example, the easy axis of Gd metal is also known to rotate upon warming by up to 60° [35], suggesting that optically-induced reorientation towards a transient easy axis could also occur. This and other similarities between our observations and other Gd-based magnets, suggest that optical control of AF order may be possible for other Gd systems. Since Gd is commonly employed in applications requiring its high $4f$ moment, this property may prove as useful for the design of deterministically controllable spintronic devices.

## Acknowledgements

We gratefully acknowledge the experimental support of the staff at beamlines UE56/1 (HZB), X11MA (SLS), and ID32 (ESRF). This work received funding from the DFG within the Emmy Noether program under Grant No. RE 3977/1, within the Transregio TRR 227 Ultrafast Spin Dynamics (Project A09) and within grant No. KR3831/5-1. We acknowledge financial support from the Spanish Ministry of Economy (MAT-2017-88374-P).

## Author contributions

L.R. and D.V.V. planned the experiment. K.Kl. and C.K. grew the crystals. Equilibrium RXD experiments were performed by Y.W.W., K.K., U.S. and L.R, and analysed by Y.W.W. and K.K. First-principles calculations were done by A.E. Time-resolved RXD experiments were performed by Y.W.W., L.R., C.S.L. and N.P., and analysed by Y.W.W. and L.R. Interpretation was done by Y.W.W., L.R., A.E. and E.V.C. The manuscript was written by Y.W.W. and L.R. All authors contributed to discussion and revision of the manuscript to its final version.

# Methods

Sample Preparation

Samples were cleaved single crystals of $GdRh_2Si_2$, grown as described in[13]. Due to the layered crystal structure, the cleaved sample face normal is precisely parallel to the tetragonal [001] axis. The crystals used were approximately 1 $mm^3$ in size, with faces much larger than the beam sizes.

Resonant X-ray diffraction (RXD)

All experiments were conducted by fulfilling the Bragg condition for the (001) magnetic reflection using incoming photon energies near the Gd $M_5$ absorption edge (~1190 eV). All experiments were conducted with $\sigma$-polarized incoming light. Correspondingly, Eq. 1 considers only the $\sigma \rightarrow \pi'$ channel ($\sigma \rightarrow \sigma'$ is 0 by symmetry for magnetic E1-E1 events). From the shape of the (001) reflection around the $M_5$ edge, we estimate an effective X-ray probe depth of 26 nm (see supplementary methods 1). The angle $\Psi$ is 0° when the $[100]$ direction is in the scattering plane.

Equilibrium RXD experiments

Experiments were conducted at beamline X11MA of the Swiss Light Source [36], using the RESOXS end station [37] and at beamline ID32 at the ESRF[38].

Following Ref. [39], in Fig. 1c the temperature dependence of $m(T)$ was fit to a mean field approximation of the form

$$\left(\frac{m}{m_0} - a\right)^2 = a^2 + \frac{1-2a}{1-2b}\left(1 - \frac{T}{T_N}\right)\left(1 - 2d + \left(\frac{T}{T_N}\right)^c\right)$$

$$c = 1 + \frac{1}{4S(S+1)} \quad . \qquad \text{(M1)}$$

The tabulated parameters for S=7/2 are $a = -1.478 \times 10^{-3}$ and $b = 2.7221 \times 10^{-2}$. The temperature dependence of $\varphi(T)$ was fit to an error function of the form:

$$\varphi(T) = A \frac{2}{\sqrt{\pi}} \int_0^T e^{-t^2} dt + B \quad . \qquad \text{(M2)}$$

Time-resolved RXD experiments

Experiments were conducted in a UHV scattering chamber using ultrashort X-ray pulses from the femtoslicing facility at beamline UE56/1-ZPM [40] at the Helmholz-Zentrum Berlin. The zone plate monochromator used in this experiment provided an energy resolution of 4 eV (see supplementary methods 1).

The pump-probe experimental scheme was conducted at 3 kHz using 1.55 eV (800 nm) p-polarized pump pulses. The X-ray repetition rate is 6 kHz such that between every pumped event an unpumped signal is recorded, and no average heating was observed. The detector is an avalanche photodiode (APD). The 1190 eV and 1.55 eV pulses arrive nearly collinearly, but the APD does not collect the pump photons which are blocked by an Al foil. The 1.55 eV and 1190 eV spot profiles were set to $(166 \pm 26 \times 321 \pm 60)\mu m^2$ and $(170 \pm 66 \times 170)\mu m^2$ [41], respectively.

The detection scheme allowed for single-photon counting, such that collected intensities represent photon counts per second. As such, the error bars in Fig. 2b,c,d are taken as $\Delta I = \sqrt{I}$. Reciprocal space scans such as those in Fig. 2b and 2c were collected for every fluence at several delays, and no broadening or shifting of the peaks was observed. Therefore only the peak heights were collected in time traces.

The time resolved experiments were conducted at four azimuths at 45° intervals: $\Psi = 15°$, 60°, 105° and 150°. The probed areas on the surface were chosen to ensure a single-domain state: two dimensional scans of the sample surface were taken before and after excitation. Chosen areas were large and homogenous, and exhibited a uniform response to excitation. The areas chosen for different azimuths were also consistent with each other, in agreement with the azimuthal trend they should exhibit according to Eq. 1. This was the case for all except the last azimuth, $\Psi = 15°$. For that dataset a large degree of domain mixing was apparent in the data, such that a mixture of $105°$ and $15°$ domains (90° apart) contributes to the observed intensity. The non-mixed signal for $\Psi = 15°$ was extracted as described in supplementary methods 2. Contamination between the 60° and 150° azimuths is also possible, but due to their large difference in intensity (see Fig. 2b,c) any appreciable effect would only occur for $\Psi = 150°$. To account for this, a small fraction (2-4%) of the $\Psi = 60°$ signal is subtracted from the 150° signal before fitting Eq. 1.

The demagnetizations in the curves in Fig. 3a were fit to an equation of the form

$$m(t)/m_0 = 1 - d_{fast}\left(1 - e^{-t/\tau_{fast}}\right) + d_{slow}\left(1 - e^{-t/\tau_{slow}}\right) \ . \quad \text{(M3)}$$

The temporal resolution is estimated at 120 fs. To account for this, fits to Eq. (M3) were conducted with a convolved Gaussian response function of 120 fs width.

Lastly, fits to Eq. 4 are done with a constant damping time of $\tau_{damp} = 150$ ps, as this does not affect the results appreciably. Fitted Values of $\zeta$ range between 116° and 97°.

Calculation of absorbed fluence:

All fluences reported are total absorbed fluences. In order to determine the optical constants of $GdRh_2Si_2$, we conducted reflectivity measurements using 1.55 eV light. From this we estimate the complex index of refraction at this photon energy as $n = n_0 + ik = (3.34 \pm 0.10) + (3.19 \pm 0.19)i$. Using this, recorded incident fluences were corrected for reflection and refraction to produce the total absorbed fluence. Note that a value of 1 mJ/cm$^2$ corresponds to an excitation density of 420 J/cm$^3$ within the first layer of the material. From this index of refraction we estimate a penetration depth of 20.6 nm at the Bragg angle $\theta$.

Anisotropy potential

The anisotropy values K presented in Fig. 4*f* were calculated from the observed oscillations. We first convert them to eigenfrequencies by correcting for the damping as $\omega_0^2 = \omega^2 + \tau_{damp}^{-2}$. We then estimate the anisotropy field $B_A$ by following the derivation in Ref. [42] (similarly to Ref. [31]) as $\omega_0 = 2\gamma B_A$, with $\gamma = g\mu_B/h$ and $g = 2$ for Gd, in which $\mu_B$ is Bohr's magneton and $h$ is Planck's constant. Finally, for a simple estimate of the constant $K$, we use the relation $B_A = K/m$, in which $m$ is the total Gd 4f moment ($\mu_B g_J J$,) and $g_J$ is the Landé factor.

First-principles calculations

The calculations were performed using a first-principles Green's function method within the multiple scattering theory[43,44] and within the density function theory in a local density approximation[45]. To correctly describe strongly localized Gd f-states, a first-principles Hubbard model with U=3eV was used[46].

The effect of temperature was simulated as a spin disorder using the disordered local moment approximation (DLM)[47,48], implemented using a coherent-potential approximation[49,50]. In the DLM approach, the absence of spin disorder corresponds to the ground state at T=0, which is the antiferromagnetic order in the case of $GdRh_2Si_2$. An increase of temperature leads to disordering of magnetic moment orientations. A fully disordered moment configuration corresponds to the paramagnetic state high above $T_N$.

The easy axis direction was calculated as follows. For a given 4f spin disorder level, the spins were aligned (anti)parallel to a given direction in the ab plane. The energy of each spin direction was calculated, and the minimal energy is taken as the easy axis angle for the given spin disorder.

## References

1. Jungwirth, T. *et al.* The multiple directions of antiferromagnetic spintronics. *Nat. Phys.* **14**, 200–203 (2018).
2. Baltz, V. *et al.* Antiferromagnetic spintronics. *Rev. Mod. Phys.* **90**, 015005 (2018).
3. Gomonay, O., Baltz, V., Brataas, A. & Tserkovnyak, Y. Antiferromagnetic spin textures and dynamics. *Nat. Phys.* **14**, 213–216 (2018).
4. Šmejkal, L., Mokrousov, Y., Yan, B. & MacDonald, A. H. Topological antiferromagnetic spintronics. *Nat. Phys.* **14**, 242–251 (2018).
5. Schlauderer, S. *et al.* Temporal and spectral fingerprints of ultrafast all-coherent spin switching. *Nature* **569**, 383–387 (2019).
6. Bhattacharjee, N. *et al.* Néel Spin-Orbit Torque Driven Antiferromagnetic Resonance in Mn2Au Probed by Time-Domain THz Spectroscopy. *Phys. Rev. Lett.* **120**, 237201 (2018).
7. Kubacka, T. *et al.* Large-Amplitude Spin Dynamics Driven by a THz Pulse in Resonance with an Electromagnon. *Science.* **343**, 1333–1336 (2014).
8. Kampfrath, T. *et al.* Coherent terahertz control of antiferromagnetic spin waves. *Nat. Photonics* **5**, 31–34 (2011).
9. Radu, I. *et al.* Transient ferromagnetic-like state mediating ultrafast reversal of antiferromagnetically coupled spins. *Nature* **472**, 205–209 (2011).
10. Stupakiewicz, A., Szerenos, K., Afanasiev, D., Kirilyuk, A. & Kimel, A. V. Ultrafast nonthermal photo-magnetic recording in a transparent medium. *Nature* **542**, 71–74 (2017).
11. Železný, J., Wadley, P., Olejník, K., Hoffmann, A. & Ohno, H. Spin transport and spin torque in antiferromagnetic devices. *Nat. Phys.* **14**, 220–228 (2018).
12. Kliemt, K. *et al.* GdRh2Si2: An exemplary tetragonal system for antiferromagnetic order with weak in-plane anisotropy. *Phys. Rev. B* **95**, 134403 (2017).
13. Kliemt, K. & Krellner, C. Single crystal growth and characterization of GdRh2Si2. *J. Cryst. Growth* **419**, 37–41 (2015).
14. Schüßler-Langeheine, C. *et al.* Resonant magnetic X-ray scattering from ultrathin Ho-metal films down to a few atomic layers. *J. Electron Spectros. Relat. Phenomena* **114–116**, 953–957 (2001).
15. Fink, J., Schierle, E., Weschke, E. & Geck, J. Resonant elastic soft x-ray scattering. *Rep. Prog. Phys.* **76**, 056502 (2013).
16. J. P. Hannon, G. T. Trammell, M. Blume, D. G. X-Ray Resonance Exchange Scattering. *Phys. Rev. Lett.* **61**, 2644 (1988).
17. Hill, J. P. & Mcmorrow, D. F. X-ray resonant exchange scattering: Polarization dependence and correlation functions. *Acta Crystallogr. Sect. A Found. Crystallogr.* **52**, 236–244 (1996).
18. Colarieti-Tosti, M. *et al.* Origin of magnetic anisotropy of gd metal. *Phys. Rev. Lett.* **91**, 157201 (2003).
19. Jensen, J. & Mackintosh, a. R. R. Rare earth magnetism: structures and excitations.

*Physics (College. Park. Md).* 403 (1991).

20. Kondorsky, E. A Review of the Theory of Magnetic Anisotropy in Ni. *IEEE Trans. Magn.* **10**, 132–136 (1974).

21. Kirilyuk, A., Kimel, A. V. & Rasing, T. Ultrafast optical manipulation of magnetic order. *Rev. Mod. Phys.* **82**, 2731–2784 (2010).

22. Thielemann-Kühn, N. *et al.* Ultrafast and Energy-Efficient Quenching of Spin Order: Antiferromagnetism Beats Ferromagnetism. *Phys. Rev. Lett.* **119**, 197202 (2017).

23. Rettig, L. *et al.* Itinerant and Localized Magnetization Dynamics in Antiferromagnetic Ho. *Phys. Rev. Lett.* **116**, 257202 (2016).

24. Wietstruk, M. *et al.* Hot-electron-driven enhancement of spin-lattice coupling in Gd and Tb 4f ferromagnets observed by femtosecond x-ray magnetic circular dichroism. *Phys. Rev. Lett.* **106**, 127401 (2011).

25. Eschenlohr, A. *et al.* Role of spin-lattice coupling in the ultrafast demagnetization of Gd1-xTbx alloys. *Phys. Rev. B - Condens. Matter Mater. Phys.* **89**, 214423 (2014).

26. Koopmans, B. *et al.* Explaining the paradoxical diversity of ultrafast laser-induced demagnetization. *Nat. Mater.* **9**, 259–265 (2010).

27. Roth, T. *et al.* Temperature dependence of laser-induced demagnetization in Ni: A key for identifying the underlying mechanism. *Phys. Rev. X* **2**, 021006 (2012).

28. Frietsch, B. *et al.* Disparate ultrafast dynamics of itinerant and localized magnetic moments in gadolinium metal. *Nat. Commun.* **6**, 8262 (2015).

29. Atxitia, U. & Chubykalo-Fesenko, O. Ultrafast magnetization dynamics rates within the Landau-Lifshitz-Bloch model. *Phys. Rev. B - Condens. Matter Mater. Phys.* **84**, 144414 (2011).

30. Zeiger, H. . *et al.* Theory for displacive excitation. *Phys. Rev. B* **45**, 768–778 (1992).

31. Sichelschmidt, J., Kliemt, K., Hofmann-Kliemt, M. & Krellner, C. Weak magnetic anisotropy in GdRh2Si2 studied by magnetic resonance. *Phys. Rev. B* **97**, 214424 (2018).

32. Brorson, S. D. *et al.* Femtosecond room-temperature measurement of the electron-phonon coupling constant γ in metallic superconductors. *Phys. Rev. Lett.* **64**, 2172–2175 (1990).

33. Low, G. G., Okazaki, A., Stevenson, R. W. H. & Turberfield, K. C. A Measurement of Spin-Wave Dispersion in MnF 2 at 4.2°K. *J. Appl. Phys.* **35**, 998 (1964).

34. Olejník, K. *et al.* Terahertz electrical writing speed in an antiferromagnetic memory. *Sci. Adv.* **4**, 3 (2018).

35. Corner, W. D. & Tanner, B. K. The easy direction of magnetization in gadolinium. *J. Phys. C Solid State Phys.* **9**, 627–633 (1975).

36. Flechsig, U. *et al.* Performance measurements at the SLS SIM beamline. *AIP Conf. Proc.* **1234**, 319–322 (2010).

37. Staub, U. *et al.* Polarization analysis in soft X-ray diffraction to study magnetic and orbital ordering. *J. Synchrotron Radiat.* **15**, 469–76 (2008).

38. Brookes, N. B. *et al.* The beamline ID32 at the ESRF for soft X-ray high energy

resolution resonant inelastic X-ray scattering and polarisation dependent X-ray absorption spectroscopy. *Nucl. Instruments Methods Phys. Res. Sect. A Accel. Spectrometers, Detect. Assoc. Equip.* **903**, 175–192 (2018).

39. Aharoni, A. *lntroduction to the Theory of Ferromagnetism*. (Oxford Iniversity Press, 1996).

40. Pontius, N., Holldack, K., Schüßler-Langeheine, C., Kachel, T. & Mitzner, R. The FemtoSpeX facility at BESSY II. *J. large-scale Res. Facil. JLSRF* **2**, A46 (2016).

41. Schick, D. *et al.* Analysis of the halo background in femtosecond slicing experiments. *J. Synchrotron Radiat.* **23**, 700–711 (2016).

42. Gurevich, A. G. & G. A. Melkov. *Magnetization Oscillations and Waves*. (CRC Press, Boca Raton, 1996).

43. Gyorffy, B. L. & Stott, M. J. Theory of Soft X-Ray Emission from Alloys. in *Proc. of the Int. Conf. on Band Structure and Spectroscopy of Metals and Alloys* (eds. Fabian, D. J. & Watson, L. M.) 385–402 (London and NY: Acad. Press, 1973).

44. Geilhufe, M. *et al.* Numerical solution of the relativistic single-site scattering problem for the Coulomb and the Mathieu potential. *J. Phys. Condens. Matter* **27**, 435202 (2015).

45. Perdew, J. P. *et al.* Atoms, molecules, solids, and surfaces: Applications of the generalized gradient approximation for exchange and correlation. *Phys. Rev. B* **46**, 6671–6687 (1992).

46. Anisimov, V. I., Zaanen, J. & Andersen, O. K. Band theory and Mott insulators: Hubbard U instead of Stoner I. *Phys. Rev. B* **44**, 943–954 (1991).

47. Gyorffy B., L., Pindor A., J., Staunton, J., Stocks G., M. & Winter, H. A first-principles theory of ferromagnetic phase transitions in metals. *J. Phys. F Met. Phys.* **15**, 1337 (1985).

48. Staunton, J., Gyorffy, B. L., Pindor, A. J., Stocks, G. M. & Winter, H. Electronic structure of metallic ferromagnets above the Curie temperature. *J. Phys. F Met. Phys.* **15**, 1387 (1985).

49. Soven, P. Coherent-Potential Model of Substitutional Disordered Alloys. *Phys. Rev.* **156**, 809–813 (1967).

50. Gyorffy, B. L. Coherent-Potential Approximation for a Nonoverlapping-Muffin-Tin-Potential Model of Random Substitutional Alloys. *Phys. Rev. B* **5**, 2382–2384 (1972).